\documentclass{aip-cp}
\usepackage[numbers]{natbib}
\usepackage{rotating}
\usepackage{graphicx}
\usepackage{upgreek}
\usepackage{lineno}
\usepackage{subfig}
\makeatletter 
\renewcommand\@biblabel[1]{#1.} 
\makeatother

\begin{document}

% Title portion
\title{Slow Extraction of Charged Ion Pulses from the REXEBIS}

\author[aff1]{N. Bidault\corref{cor1}}
\author[aff1]{J. A. Rodriguez}
\author[aff1]{M. Lozano}
\author[aff1]{S. Sadovich}

\affil[aff1]{CERN, CH-1211 Geneva, Switzerland}
%\affil[aff2]{The Joint Institute for Power and Nuclear Research Sosny, BY-220109 Minsk, Belarus}
\corresp[cor1]{Corresponding author: niels.killian.noal.bidault@cern.ch}

\maketitle

\begin{abstract}
The Isotope mass Separator On-Line DEvice (ISOLDE) facility located at CERN, produces and transports Radioactive Ion Beams (RIBs) at low or high energy through the REX/HIE-ISOLDE linear accelerator, for nuclear physics, astrophysics, solid-state physics and applied-physics purposes. Increasing the charge state of the ions is a prerequisite for efficient acceleration and is accomplished by an Electron Beam Ion Source (REXEBIS). For more effective event discrimination at the experimental detectors, such as the MINIBALL spectrometer, it is advantageous to increase the pulse width of extracted ions from this EBIS. A Slow Extraction scheme is presented which uses a function comprised of discrete voltage steps to apply the extraction potential to the EBIS trap barrier. This function effectively stretches the pulse length of both stable and radioactive ion beams, with different mass-to-charge ratios and provides for extracted pulse widths in the millisecond range. Key operational parameters of the EBIS impacting the average ionic temperature and its axial energy spread are discussed, in order to anticipate changes in the resulting ion pulse time structures during experimental runs.
\end{abstract}

% Head 1
\section{INTRODUCTION}
Prior to re-acceleration, the ion beam from the ISOL target is accumulated and cooled in a gas-filled Penning trap (REXTRAP) and then charge-bred in the REXEBIS \cite{bib1}. The ensemble of REX/HIE-ISOLDE sequences is cycled according to the repetition rate of the machine. In order to increase the detection efficiency at the experimental stations, the pulse width of the ion beam extracted from the REXEBIS should be lengthened. The typical extraction method which consists of quickly reducing the potential on the barrier trapping the ions, does not assure sufficient temporal event discrimination due to the high linear current density of ions within the brief ($\leqslant$ 200 $\upmu$s) extracted ion pulse. Consequently, we have developed the Slow Extraction method at REXEBIS to achieve an ion pulse length in the millisecond range. The concept of ion pulse stretching has already been used extensively \cite{bib2} and theorized \cite{bib3}.

\section{COMMISSIONING OF THE ION PULSE SHAPING}
Inside of the REXEBIS, ions are radially trapped and ionized by an electron beam to meet a mass-to-charge ratio (A/q) within the acceptance of the REX/HIE-ISOLDE linear accelerator, between 2.5 and 4.4 \cite{bib4}. The ion trap resides within a magnetic field of 2 T, produced by a superconducting solenoid. The electron beam emitted from the LaB$_6$ cathode is usually operated with current limited to about 200 mA for lifetime reasons and has an energy equal to the potential difference between the electron cathode and the trapping region. The axial confinement of the ions is made by two 1.2 kV barriers at the edges of three 0.7 kV cylindrical drift electrodes, all coaxial (Figure 1). 

\begin{figure}[h]
	\centerline{\includegraphics[height=460pt, trim={430pt 0pt 10pt 0pt},angle = -90]{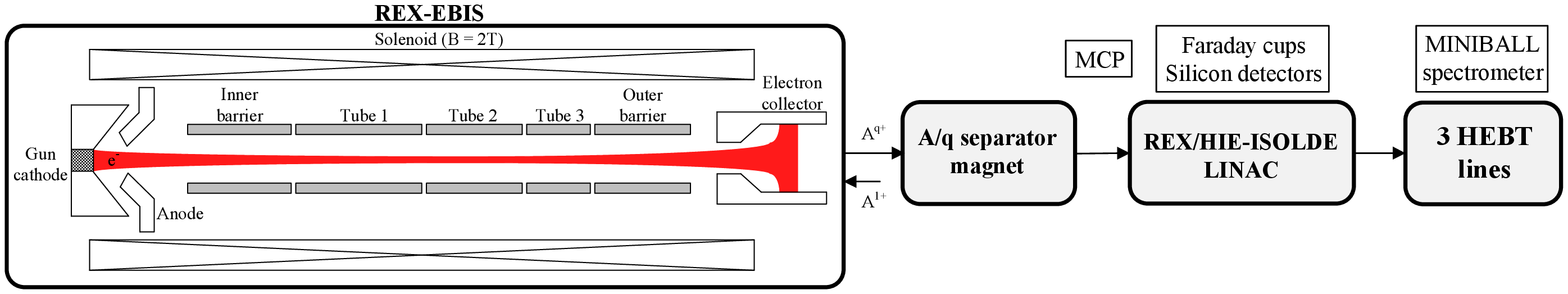}}
	\caption{Schematic of the REXEBIS, blocks of the main elements and diagnostics along the experimental set-up}
\end{figure}

Once a sufficient population of ions reaches the required charge-state, all ions are extracted from the REXEBIS by quickly reducing the extraction barrier potential. After extraction, the highly charged ions of interests are selected by an A/q separator dipole. The pulse length of the Radio-Frequency (RF) systems of the REX-ISOLDE normal-conducting linear accelerator is constrained by a maximum duty cycle of 10\% and by the average power limitation of the 9-gap IH resonator (2.5 kW). Typically, when operated at 50 Hz repetition rate, the RF pulse could be up to 2 ms long and, ideally, the pulse width of the REXEBIS extracted ion beam should be the same.

The MINIBALL spectrometer at the end of the first out of three High Energy Beam Transfer (HEBT) lines can record the time-of-flight distribution of the incoming ions \cite{bib5}. In addition, two other types of diagnostic along the REX/HIE-ISOLDE accelerating line are used to develop the Slow Extraction method at REXEBIS. A Phosphor-screen Micro-Channel Plate (MCP) is installed following the A/q separator dipole and offers a good approximation of the envelope of the time structure for ion pulses with intensities above 10 pA. In the case of very low intensity ion beam, five solid-state Silicon detectors provide digital counting of every ion event signal within a time resolution of 1 $\upmu$s, allowing one to reconstruct a histogram of the extracted ion pulse \cite{bib6}.

The first step to determine the extraction barrier voltage temporal function that will stretch the extracted ion pulse to the millisecond range, is to measure the axial energy distribution of the ion beam extracted from the REXEBIS and transported to the REX/HIE-ISOLDE linear accelerator. For this measurement, the extraction barrier is reduced and held for 2 ms at successively lower voltage ``thresholds'' during consecutive EBIS extraction cycles. After the 2 ms ``threshold extraction'' the barrier is lowered to completely empty the EBIS ion trap in preparation for the next ion breeding cycle. During the first 2 ms, only the ions having an axial energy above the threshold barrier voltage are able to escape and to consecutively be captured by the RF pulse of the RFQ. The population of ions with an axial energy under the threshold voltage is extracted too late to be accelerated through the linear accelerator and is lost. By progressively lowering the threshold voltage from 1.2 kV to 0 kV, while acquiring with a Faraday cup the intensity of the ion beam passing the RFQ, one is able to reconstruct the  axial energy distribution of the ions de facto transported through the accelerator. For very low intensity beam not measurable with a Faraday cup, the method has been validated using a Silicon detector. The extraction barrier voltage is controlled by a step-function, with a minimum time interval of $\Delta t = 10$ $\upmu$s, therefore the discrete integral of the axial energy distribution $F(U)$ is set equal to a fraction of the total amount of ions per pulse. Considering the required length of the ion pulse $T_p$ and the initial barrier voltage $V_{t_0} = 1.2$ kV, all successive $V_{t_i}$ values of the step-function can be numerically deduced from:
\begin{eqnarray}
\int_{V_{t_i}}^{V_{t_{i+1}}} \! F(U) \, \mathrm{d}U&=&\frac{\Delta t}{T_p}\int_{0}^{V_{t_0}} \! F(U) \, \mathrm{d}U
\end{eqnarray}
An example of the resulting pulse shape elongated to the maximum possible length of the RF pulses in June 2017 (0.7 ms), for ions with A/q = 4.12 is presented in Fig. 2. The time structure displayed, acquired with a Silicon detector and the MINIBALL spectrometer, is directly obtained from the application of Equation 1, without refinement nor smoothing of the barrier voltage function. The ripples in the beginning of the ion pulse distribution are under investigation, they were not necessarily present after each pulse shaping commissioning and may be a consequence of the different ramping voltages applied to the drift electrodes during extraction. By comparison, the ion pulse width deriving from an abrupt opening of the extraction barrier is less than 200 $\upmu$s long and is essentially dependent on the trap length, on the ion mass, on the energy spread and on the slew rate of the extraction electrode (55 V/$\upmu$s).
\begin{figure}[h]
	\centering
	\begin{tabular}[b]{c}
		\includegraphics[height=90pt,trim={20pt 20pt 20pt 20pt}]{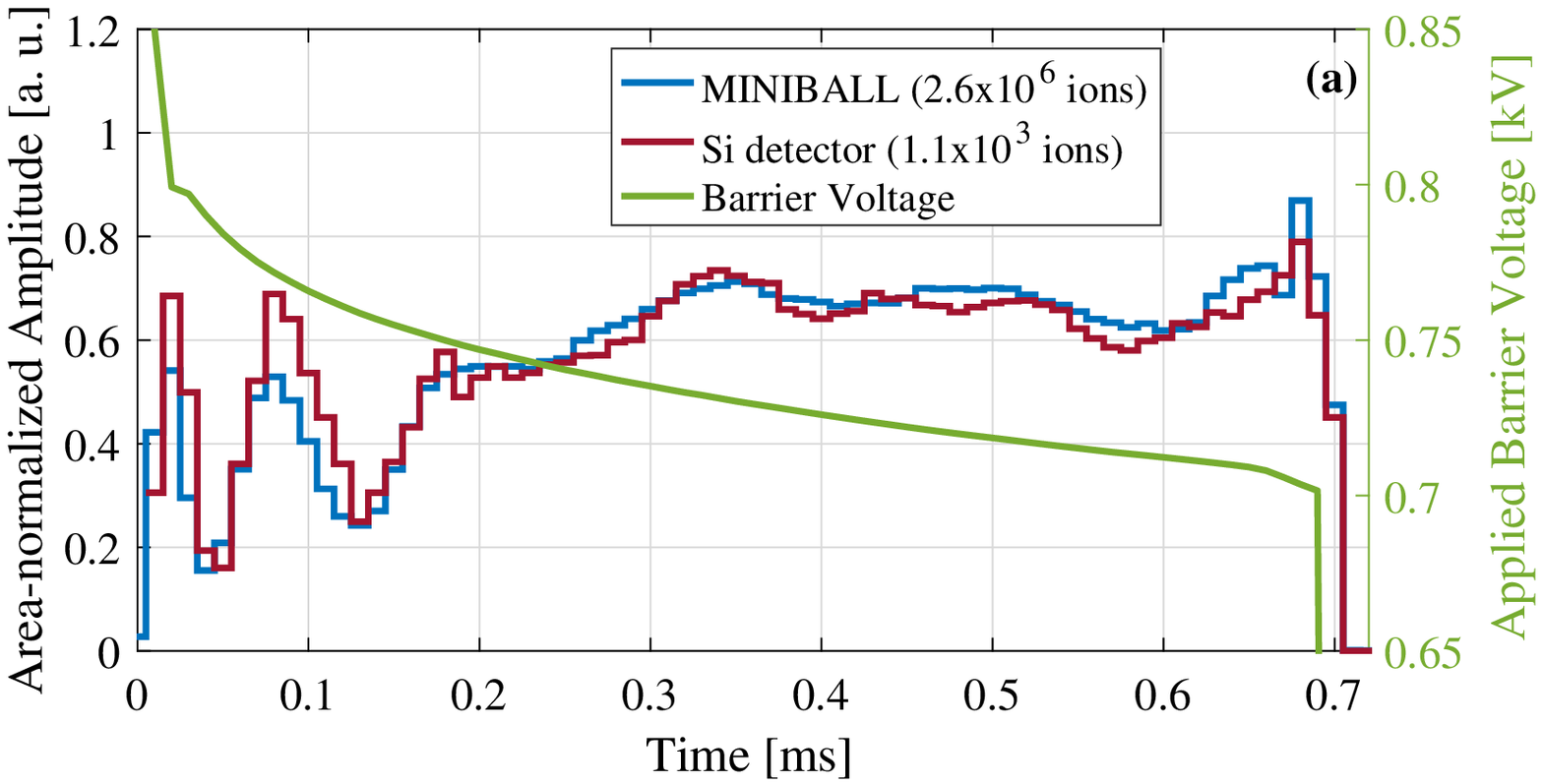} \\
	\end{tabular} \qquad
	\begin{tabular}[b]{c}
		\includegraphics[height=90pt, trim={20pt 20pt 20pt 20pt}]{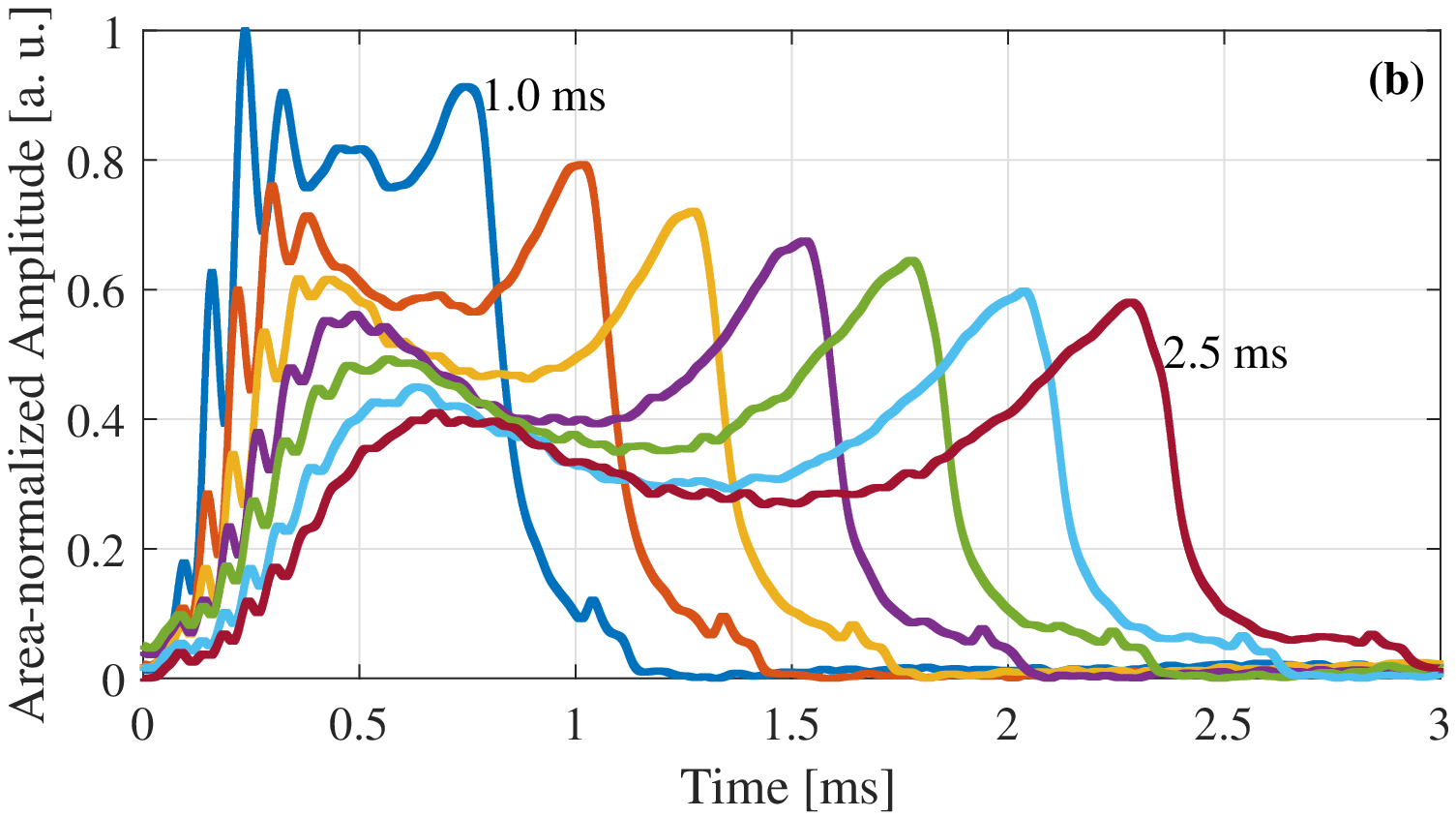} \\
	\end{tabular}
	\caption{(a) Slow Extraction of $^{140}$Sm$^{34+}$ time structure. (b) $^{132}$Xe$^{32+}$ resulting time structures lengthening, acquired with the MCP, when extending the extraction voltage step-function from 1.0 ms to 2.5 ms (0.25 ms steps).}
\end{figure}

Since the commissioning of the new Slow Extraction method in March 2016, time structure shaping from 0.7 ms to 5 ms has been achieved for more than thirty different radioactive or stable ion beams, satisfying the A/q range requirement. The possibility to adapt the ion pulse to any desired length has also been demonstrated and will be useful when longer RF pulse lengths are available.
 
\section{REXEBIS OPERATIONAL PARAMETERS INFLUENCE ON THE ION PULSE SHAPE}
After having commissioned the Slow Extraction, several operational parameters of the REXEBIS still might be adjusted during experimental runs and it becomes necessary to anticipate their potential influences on the extracted ion pulse shape. The trapping time may be adjusted for charge breeding of different radioactive isotopes under study. Furthermore, the electron beam current and energy may be reduced to extend the cathode lifetime. Variation of the electron beam energy within the typical operating range of 3 to 4 keV, were not measured to have a significant influence on the ion pulse shape. However, the trapping time and the electron beam current play a role on the ion axial energy distribution and will result in altered ion pulse shapes for a given extraction voltage modulation (Fig. 3).
\begin{figure}[h]
	\centering
	\begin{tabular}[b]{c}
		\includegraphics[height=90pt,trim={20pt 20pt 20pt 20pt}]{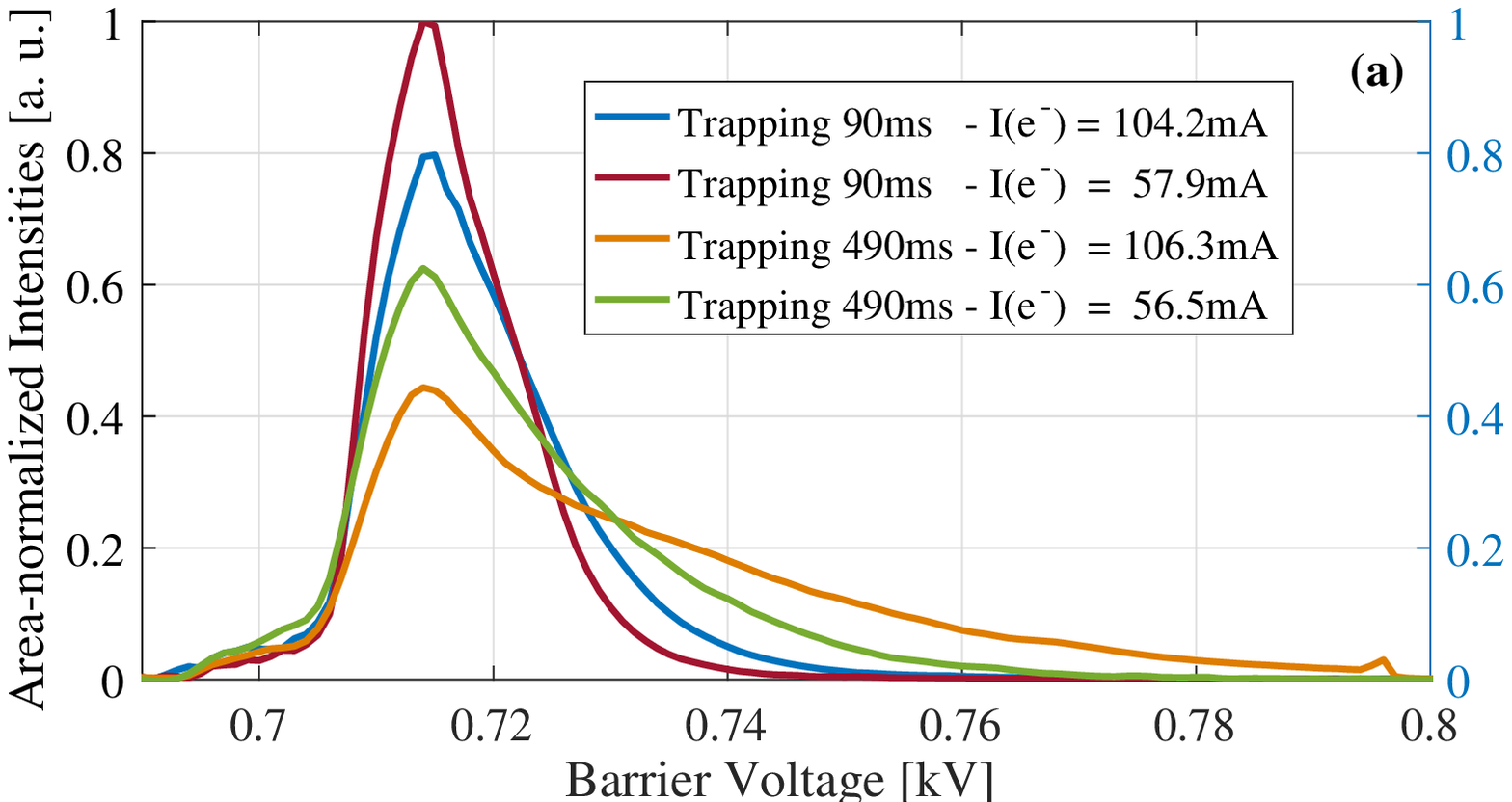} \\
	\end{tabular} \qquad
	\begin{tabular}[b]{c}
		\includegraphics[height=90pt, trim={20pt 20pt 20pt 20pt}]{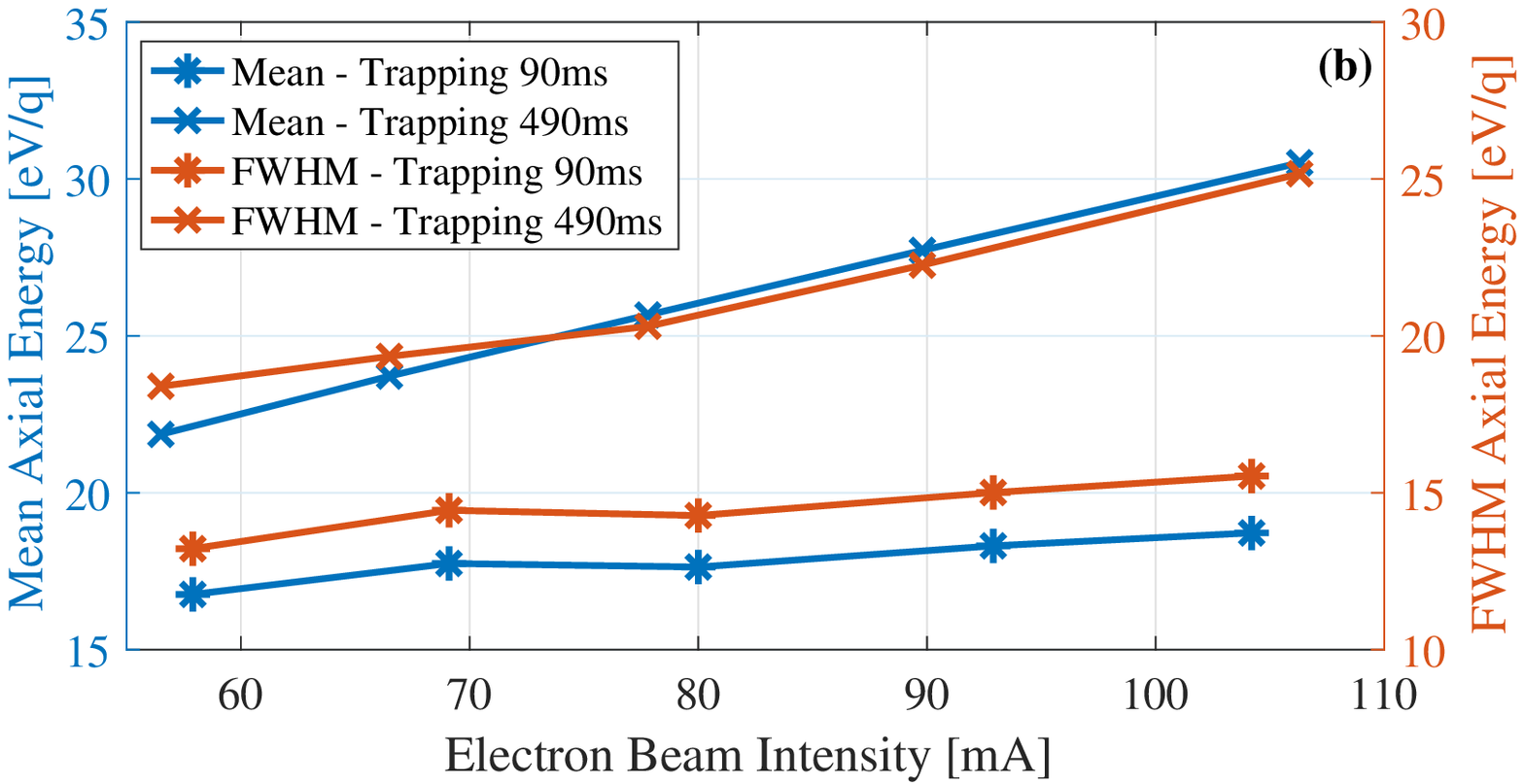} \\
	\end{tabular}
	\caption{(a) Ion axial energy distribution of $^{14}$N$^{5+}$ measured for two different trapping times and electron beam intensities. (b) Influence of the electron beam intensity on $^{14}$N$^{5+}$ axial mean energy and spread}
\end{figure}

As the trapping time increases, the neutralization of the electron beam by ions rises, this will decrease the radial potential and result in an increase in the mean extracted ion energy. The ion heating rate from long-range Coulomb elastic impacts, expressed by L. Landau and E. L. Spitzer, describes a relation of proportionality between the ionic temperature and the electron beam current density \cite{bib7}. A calculation of the electron beam radius via Herrmann theory suggests a variation of only 0.02\% in the range of electron currents used, thus it was considered constant. By fitting the mean axial energy as a linear function of the electron beam intensity, one obtains a ratio of 4.50 between the proportionality coefficients of the two slopes, with 490 ms and 90 ms trapping times. This ratio is twice smaller than the trapping time increase multiplied by the raise of $^{14}$N$^{5+}$ ions extracted, notably indicating that radial escape occurs during the confinement. The mean axial energy increase as a function of the trapping time is not linear, it will reach a plateau of equilibrium after a certain time depending upon the electron beam density and the evaporative cooling.
% Sections that will go in second font
\section{CONCLUSION}
A Slow Extraction method was developed and implemented at REXEBIS, for a variety of ion beams in the A/q range from 2.5 to 4.4 and with a time structure length in the millisecond range. The Slow Extraction scheme reduces saturation issues at the experimental stations. Two essential operational parameters, the electron beam current and the ion trapping time, were identified as influencing the ionic temperature and the ion axial energy spread. This knowledge will allow us to anticipate changes and adjust the slow extracted ion pulse shape accordingly to maintain a uniform counting rate at the detectors.
 
% References

\nocite{*}
\bibliographystyle{aipnum-cp}%
\bibliography{SlowExtraction}%

\end{document}